\documentstyle[12pt,aaspp4]{article}
\begin{document}
\parindent=1.0cm

\title{High Angular Resolution $JHK$ Imaging of the Centers of the Metal-Poor 
Globular Clusters NGC5272 (M3), NGC6205 (M13), NGC6287, and NGC6341 (M92)} 

\author{T. J. Davidge \altaffilmark{1} \& S. Courteau \altaffilmark{1}}

\affil{Herzberg Institute of Astrophysics, \\ National Research Council of 
Canada,\\ 5071 W. Saanich Road, Victoria, BC Canada V8X 4M6\\ 
{\it email: tim.davidge@hia.nrc.ca, stephane.courteau@hia.nrc.ca}}

\altaffiltext{1}{Visiting Astronomer, Canada-France-Hawaii Telescope, which 
is operated by the National Research Council of Canada, the Centre National de 
la Recherche Scientifique, and the University of Hawaii}

\begin{abstract}

	The Canada-France-Hawaii Telescope (CFHT) Adaptive Optics Bonnette 
(AOB) has been used to obtain high angular resolution $JHK$ images of the 
centers of the metal-poor globular clusters NGC5272 (M3), NGC6205 (M13), 
NGC6287, and NGC6341 (M92). The color-magnitude diagrams (CMDs) derived from 
these data include the upper main sequence and most of the red giant branch 
(RGB), and the cluster sequences agree with published photometric measurements 
of bright stars in these clusters. The photometric accuracy is limited by PSF 
variations, which introduce systematic errors of a few hundredths of a 
magnitude near the reference star. The NGC6287 CMDs are of particular interest, 
as they are free of the scatter caused by foreground stars and differential 
reddening that have plagued previous efforts to study the stellar content 
of this inner spheroid cluster. The horizontal branch (HB) of NGC6287 
contains stars spanning a broader range of temperatures than that of NGC6341. 
NGC6287 is thus the only known member of the metal-poor inner spheroid to 
deviate from the HB morphology versus metallicity relation defined by other 
clusters in the inner Galaxy. 

	The clusters are paired according to metallicity, and the near-infrared 
CMDs and luminosity functions are used to investigate the relative ages within 
each pair. The near-infrared CMDs provide the tightest constraints on the 
relative ages of the classical second parameter pair NGC5272 and NGC6205, and 
indicate that these clusters have ages that differ by no more than $\pm 1$ Gyr. 
These results thus support the notion that age is not the second parameter. We 
tentatively conclude that NGC6287 and NGC6341 have ages that differ by no 
more than $\pm 2$ Gyr. However, the near-infrared spectral energy distributions 
of stars in NGC6287 appear to differ from those of stars in outer 
halo clusters, bringing into question the validity of this age estimate.

\vspace{0.3cm}

\end{abstract}

\section{INTRODUCTION}

	Deep photometric studies of globular clusters have traditionally been 
restricted to the less crowded outer regions of these objects. While surveys of 
this nature with electronic detectors have permitted the properties of 
relatively faint main sequence stars in nearby clusters to be studied in 
detail, it has been difficult to assemble photometry of statistically 
significant samples of stars in the more advanced stages of evolution using the 
same instrumentation. The extent of this problem can be reduced by observing 
the central regions of clusters, where the high stellar densities simplify 
efforts to obtain photometry of stars spanning 
a range of brightnesses and evolutionary phases, including the brightest 
portions of the red giant branch (RGB). Studies of the central regions of 
clusters also reduce the effects of foreground star contamination 
and, especially when conducted at infrared wavelengths, 
differential reddening, both of which frustrate efforts to probe the nature 
of low latitude clusters -- objects that will provide important clues 
into the early evolution of the inner Galaxy. The cost of observing stars 
in these high-density environments is increased crowding, and angular 
resolutions approaching the diffraction limits of moderately large telescopes 
are required to survey all but the brightest stars.

	In the current paper, high angular resolution 
near-infrared images obtained with the Canada-France-Hawaii Telescope 
(CFHT) Adaptive Optics Bonnette (AOB) are used to probe the central stellar 
contents in two metal-poor cluster pairs: NGC5272 (M3) $+$ NGC6205 (M13) and 
NGC6287 $+$ NGC6341 (M92). The clusters in each pair were selected to have 
similar metallicities and distances to facilitate differential 
photometric comparisons. The basic properties of the target clusters, including 
[Fe/H], $E(B-V)$, distance from the Galactic Center, R$_{GC}$, the $V$ 
brightness of the horizontal branch (HB) corrected for reddening, $V^{HB}_0$, 
the central surface brightness in $V$ corrected for reddening, $\mu^{V}_{0}$, 
and the integrated cluster brightness, M$_V$, are summarized in Table 1. The 
data in this Table were taken from the May 15, 1997 version of the Harris 
(1996) database; the reddening corrections assume that A$_{V}/E(B-V) = 3.1$.

	The clusters NGC5272 and NGC6205, which are a classic second-parameter 
pair, have been the subject of numerous photometric and spectroscopic 
investigations at visible wavelengths. However, only a modest body of infrared 
photometric data has previously been obtained for these clusters. Cohen, 
Frogel, \& Persson (1978) recorded $JHK$ aperture measurements of selected 
bright giants in the outer regions of both clusters, while Davidge \& Harris 
(1995) used moderately deep $J$ and $K$ images to investigate the 
stellar content of giants and main sequence stars in NGC6205. 

	NGC6287 and NGC6341 are among the most metal-poor clusters in 
the Galaxy. NGC6341 has been well-studied at visible wavelengths and 
is a standard benchmark for constraining the age of the 
Universe (e.g. Salaris, Degl'Innocenti, \& Weiss 1997, Bolte \& Hogan 1995). 
Cohen {\it et al.} (1978) also obtained near-infrared aperture measurements of 
bright stars in the outer regions of this cluster. NGC6287 is located 
in the inner regions of the Galaxy. The close proximity to the Galactic Center, 
size, and metallicity of NGC6287 lead van den Bergh (1993) 
to suggest that this cluster may be one of the oldest objects in the 
Galaxy. Efforts to interpret the visible (Stetson \& West 1994) and 
near-infrared (Davidge 1998) CMDs of the outer regions of NGC6287 are 
complicated by differential reddening and field star contamination. 
Davidge (1998) derived an age of 12 Gyr for NGC6287 based on the $K$ LF of 
sub-giant branch (SGB) stars and comparisons with the Bergbusch \& VandenBerg 
(1992) models. 

	The data presented in this paper will be used to investigate the 
photometric characteristics of the cluster stars and assess relative ages 
within each cluster pair. The data were recorded with the same instrumental 
configuration over a period of only two nights and include the main sequence 
turn-off (MSTO); hence, these data are well suited to the 
task of relative age determination. The observations and reductions are 
described in \S 2, while the photometric measurements, LFs and CMDs 
are discussed in \S 3. This information is used to investigate the 
relative ages of each pair of clusters in \S 4. A summary and discussion of 
the results follows in \S 5.

\section{OBSERVATIONS AND REDUCTIONS}

	The data were recorded on the nights of UT March 8/9 (NGC5272, NGC6205, 
and NGC6287) and March 9/10 (NGC6341) 1998 with the CFHT AOB (Rigaut {\it et 
al.} 1998) and KIR camera. The latter contains a $1024 \times 1024$ Hg:Cd:Te 
array with an angular scale of 0.034 arcsec per pixel, so that the total imaged 
field is $35 \times 35$ arcsec. The CFHT AOB uses natural guide stars as 
reference beacons for atmospheric compensation. These were selected at the 
telescope, and centered on the detector to reduce anisoplanaticism.

	Data were recorded through $J, H$, and $K$ filters, with a 
complete observing sequence consisting of 4 exposures 
per filter. The exception was NGC5272, for which 8 exposures per filter 
were recorded. The telescope pointing was offset following each exposure to 
define a $0.5 \times 0.5$ arcsec square dither pattern. Additional details of 
the observations, including right ascension ($\alpha$) and declination 
($\delta$), total exposure times, and the FWHM of the central peak in the 
point spread function (PSF), averaged over the entire field, are summarized in 
Table 2. 

	With the exception of NGC6205, which was observed 
during a brief episode of bad seeing, the FWHM entries for 
$H$ and $K$ are close to the telescope diffraction limit. However, 
at moderate Strehl ratios the FWHM provides an incomplete picture 
of image quality, as much of the energy in the PSF is contained 
outside of the central peak (e.g. Davidge {\it et al.} 1997ab); 
indeed, the AO-compensated PSF was traced out to a radius of at 
least 1.5 arcsec in observations of moderately bright stars recorded on the 
same nights as the cluster data. The diameter of an 
aperture containing 50\% of the PSF energy provides a means of quantifying 
the light distribution, and the 50\% encircled energy diameters 
for the NGC6287 data, which should also be representative of the NGC5272 and 
NGC6341 observations, are 0.42 ($J$), 0.25 ($H$), and 0.25 arcsec ($K$). 

	The raw images contain signatures introduced by the telescope, 
instrumentation, and sky that must be removed before making photometric 
measurements. Some of these signatures are multiplicative in nature (throughput 
variations introduced by the optics and pixel-to-pixel differences in detector 
sensitivity), while others are additive (detector bias and dark current, 
thermal emission from objects along the optical path, and sky emission). 
Calibration frames monitoring these signatures were either recorded (darks, 
flats) or derived (thermal emission) from the observations. 
Dark frames for each integration time were recorded mid-way through the run, 
while dome flats were constructed by subtracting exposures of 
the dome white spot taken with the flat-field lamps `off' from those 
recorded with the lamps `on'. This differencing procedure removes the additive 
contributions from detector dark current and thermal emission, while preserving 
the multiplicative flat-field and detector sensitivity variations. Images 
showing the thermal emission pattern for each filter were constructed by 
median-combining flat-fielded images of blank sky regions, which were recorded 
at various times throughout the run.

	The data were reduced using the following sequence: (1) dark 
subtraction, (2) division by dome flats, (3) subtraction of the DC sky level, 
which was estimated by taking the mode of the pixel 
intensity distribution in each flat-fielded frame, and (4) 
subtraction of the thermal signature. The results for each field were 
registered to correct for the offsets introduced while dithering, and 
then median-combined. The final $K$ images are shown in Figures 1 -- 4.   

\section{PHOTOMETRIC MEASUREMENTS}

	The photometric measurements were made using the PSF-fitting routine 
ALLSTAR (Stetson \& Harris 1988), which is part of the DAOPHOT (Stetson 1987) 
photometry package. Anisoplanaticism causes the PSF to vary with distance from 
the AO reference star, and the extent of these variations in the NGC6287 field 
is illustrated in Figure 5, where the $J$ and $K$ PSFs in two different annuli 
centered on the reference star are compared. The effects of anisoplanaticism 
are greatest in $J$, where the FWHM changes 
from 0.18 arcsec to 0.28 arcsec between the inner and outer annuli. 
For comparison, the FWHM changes from 0.12 and 0.14 arcsec in $K$. 

	PSF variations complicate efforts to measure accurate stellar 
brightnesses. To quantify the affects of PSF variations in the current data, 
stellar brightnesses were measured from the NGC6287 $J$ and $K$ 
images using PSFs constructed from stars (1) distributed over the entire field, 
(2) when r$_0$, the distance between a star and the reference source, is $\leq 
8.5$ arcsec, and (3) when r$_0 \geq 8.5$ arcsec.

	The brightnesses of stars with r$_0 \geq 8.5$ arcsec measured with the 
first and third set of PSFs are very similar, indicating that the use of a 
constant PSF does not significantly affect photometric measurements 
at moderately large distances from the reference star. 
The mean brightness differences among moderately bright 
stars when using these PSFs are $\overline{\Delta_{13}} = 0.01 \pm 0.03$ mag 
(J), and $0.02 \pm 0.03$ mag (K), where the quoted errors are dominated by 
uncertainties in the aperture corrections. The standard 
deviations about the $\overline{\Delta_{13}}$ values provide 
additional insight into the affects of PSF variations, and these are 
$\sigma_{13} = 0.003$ mag (J) and 0.001 mag (K). This comparison indicates 
that the use of a PSF constructed from stars over the entire field 
affects brightness measurements by $\sim 0.01$ mag when $r_0 \geq 8.5$ arcsec. 
Roughly 80\% of the stars in our sample are located in this region.

	PSF variations are of greater concern close to the reference source. 
Indeed, the mean brightness differences between measurements made with the 
first and second set of PSFs for stars with $r_0 \leq 8.5$ arcsec are 
$\overline{\Delta_{12}} = -0.07 \pm 0.14$ (J) and $-0.10 \pm 0.06$ (K). The 
errors reflect the large uncertainties in the aperture corrections for 
the measurements made using the second set of PSFs, as the number of PSF stars 
is small when $r_0 \leq 8.5$ arcsec, and there is significant crowding, as the 
reference source is located close to the cluster center. The standard 
deviation about the $\overline{\Delta_{12}}$ values are 
$\sigma_{12} = 0.03$ mag (J) and 0.01 mag (K). Hence, PSF variations affect 
brightness measurements near the reference source by a few hundredths of a 
magnitude. We emphasize that only 20\% of the stellar sample is located within 
8.5 arcsec of the reference source.

	The PSF variations introduced by anisoplanaticism depend 
on a number of environmental (e.g. wind speed and direction, altitude 
distribution of turbulence layers, location and brightness 
of the reference source etc) and instrumental (e.g. the alignment of optical 
components and the detector, order of compensation provided by the AO system 
etc) factors and, with the limited environmental information 
currently available at most telescopes, can only be modelled empirically. 
While DAOPHOT has the capability of constructing a variable PSF, a large number 
of bright, preferably isolated, stars are required, and efforts to construct 
a variable PSF that produced better photometric measurements than a single PSF, 
constructed from stars over the entire field for each cluster$+$filter 
combination, were confounded by crowding. Given these difficulties, and also 
considering that the experiments described above indicate that the 
majority of stars are not affected by the use of a constant PSF, we elected to 
use a single PSF to obtain photometric measurements for subsequent analysis.

	The photometric calibration was defined using 
UKIRT faint standard stars (Casali \& Hawarden 1992), which were observed 
at various times throughout the two night run. The standard star brightnesses 
were corrected for atmospheric extinction using the coefficients derived by 
Guarnieri, Dixon, \& Longmore (1991), and the results were then used 
to derive photometric zeropoints and first order color terms. The 
uncertainties in the photometric zeropoints infered from the scatter in the 
seven standard star measurements are: $\pm 0.010 (J)$, $\pm 0.005 (H)$, and 
$\pm 0.009$ mag $(K)$.

	The brightnesses at which sample incompleteness 
becomes significant can be estimated from the turn-over 
points at the faint end of the LFs. The $K$ LFs of these clusters are plotted 
in Figure 6, where n$_{0.5}$ is the number of stars per 0.5 mag interval per 
square arcsec. It is evident that the data are complete to 
$K = 18 - 19$ mag, which is comparable to, or slightly fainter 
than, the MSTO in each cluster.

	The $(H, J-H)$, $(K, H-K)$, and $(K, J-K)$ CMDs are plotted in Figures 
7, 8, and 9. The RGB, SGB, MSTO, and HB sequences are clearly evident in the 
$(H, J-H)$ and $(K, J-K)$ CMDs of all four clusters; an asymptotic 
giant branch (AGB) component is also apparent in the NGC5272, NGC6205, and 
NGC6287 CMDs. The NGC5272 CMDs involving the $J$ filter show the 
largest photometric scatter, as these data have unusually large PSF variations. 
However, the scatter in the other CMDs is modest, and this 
is demonstrated in Figure 10, where the $H-K$ color 
distributions of stars with $K$ between 12 and 14.5, an 
interval over which the $(K, H-K)$ CMD is practically vertical and 
contamination from the HB is small, are shown. The standard 
deviations of these distributions range from a low of $\pm 0.023$ mag for 
NGC6205 to a high of $\pm 0.046$ mag for NGC5272. If the random uncertainties 
in the $H$ and $K$ measurements are assumed to be equal, then 
these comparisons suggest that the internal photometric 
errors in the $H$ and $K$ measurements of moderately bright stars 
are $\pm 0.016$ mag for NGC6205 and $\pm 0.033$ mag for NGC5272. 
Two caveats that should be considered when assessing these 
results are that (1) the presence of AGB stars will 
broaden the distributions slightly, and (2) the PSF-variation experiments 
described earlier indicate that the widths of the color distributions 
are lower limits that apply mainly to stars with $r_0 \geq 8.5$ arcsec. 

	Comparisons with published infrared photometric measurements verify the 
color calibration of the AOB data. The open squares in Figures 7, 8, and 9 are 
the single element detector measurements published by Cohen {\it et al.} (1978) 
for bright giants in NGC5272, NGC6205, and NGC6341 and the array 
detector measurements published by Davidge (1998) of bright stars near the 
center of NGC6287. Crowding and, to a lesser extent, differential reddening 
cause the scatter in the Davidge (1998) measurements to be larger than in the 
Cohen {\it et al.} measurements, which are restricted to stars in the outer 
regions of the clusters. It is evident that the published data are in 
reasonable agreement with the AOB observations.

	The brightnesses and colors of key features on the CMDs that provide 
age and distance information at infrared wavelengths have been measured, and 
the results are listed in Table 3. The brightness of the HB is the standard 
means of judging cluster distances at visible wavelengths. However, the HB is 
not considered here since (1) it forms an almost vertical sequence on 
infrared CMDs, and (2) while relations between RR Lyrae infrared brightness and 
period have been defined (e.g. Carney, Storm, \& Jones 1992; Longmore {\it et 
al.} 1990), NGC6287 has not yet been extensively surveyed for variable stars. 
Therefore, the brightness of the RGB-tip is adopted to gauge 
relative distances. Stochastic effects may compromise RGB-tip brightness 
measurements since evolution on the upper giant branch proceeds at a 
rapid pace. However, numerical simulations indicate that only a modest 
survey of bright stellar content is required to determine the RGB-tip 
brightness to an accuracy of $\pm 0.1$ mag in a typical globular cluster 
(Crocker \& Rood 1984). 

	Stars near the RGB-tip are saturated in the AOB data. Therefore, 
the $J$, $H$ and $K$ brightnesses of the RGB-tip for NGC5272, NGC6205, 
and NGC6341 were estimated from the brightest stars in the Cohen {\it 
et al.} (1978) sample. These same Cohen {\it et al.} (1978) measurements 
were considered by Frogel {\it et al.} (1983) in their study of the bolometric 
brightness of the RGB-tip in a much larger sample of globular clusters, 
and the RGB-tip brightnesses derived for NGC5272, NGC6205, and NGC6341 
fell within the $\pm 0.1$ mag scatter envelop defined by the larger 
cluster sample, lending confidence to the RGB-tip brightness 
estimates for these three clusters. Davidge (1998) measured the $K$ 
brightness of the RGB-tip for NGC6287, and RGB-tip brightnesses in $J$ and 
$H$ were obtained from the data listed in Table 3 of that paper. The RGB-tip 
brightnesses listed in Table 3 of the current paper have an 
estimated uncertainty of $\pm 0.1$ mag.

	RGB-tip brightnesses can be used to estimate absolute distances, 
although the calibration is very uncertain. The Bergbusch \& VandenBerg (1992) 
isochrones, when transformed onto the near-infrared observational plane using 
the color relations given by Davidge \& Harris (1995a), predict that M$_K = 
-5.9$ for metal-poor clusters, so that the predicted distance moduli for the 
clusters are 15.1 (NGC5272), 14.3 (NGC6205), 14.6 (NGC6287), and 
14.8 (NGC6341). For comparison, the Bertelli {\it et al.} (1994) models 
predict an RGB-tip brightness that is 0.2 mag fainter in $K$. The resulting 
distance moduli are reasonably consistent with those predicted by the 
Carney {\it et al.} (1992) RR Lyrae calibration, which gives 
distance moduli of 14.9 (NGC5272), 14.1 (NGC6205), and 14.4 (NGC6341) 
with the cluster properties given in Table 1.

	While the brightness of the MSTO is a prime age 
indicator, it can not be measured accurately from these data as the CMDs 
are almost vertical at this point. However, the brightness of the SGB provides 
age information that is comparable to that of the MSTO brightness (Chaboyer 
{\it et al.} 1996). For this study the SGB brightness 0.1 mag redward of the 
MSTO was measured from the ridgeline of each CMD, and the estimated uncertainty 
in these measurements is $\pm 0.1$ mag. The $J-H$ and $J-K$ colors of the MSTO 
are very well-defined, and the estimated uncertainties in the observed MSTO 
colors are $\pm 0.03$ mag, 

	The de-reddened MSTO colors for each cluster, derived using the Rieke 
\& Lebofsky (1985) reddening law, are listed in Table 4. The MSTO colors for 
NGC6341 are the smallest of the group, as expected given that this is the most 
metal-poor cluster. However, the de-reddened MSTO colors of NGC6287 are 
abnormally large, especially in $J-K$. It is worth noting that the uncertainty 
in $E(B-V)$ for NGC6287 is relatively large, amounting to $0.1 - 0.2$ mag. 
$E(B-V)$ estimates for this cluster derived from data at visible wavelengths 
favour a value near 0.6 (Zinn 1980, Reed {\it et al.} 1988, Stetson \& West 
1994), while far-infrared properties of interstellar dust (Schlegel, 
Finkbeiner, \& Davis 1998) and the near-infrared colors of giant branch 
stars (Davidge 1998) favour 0.8. Nevertheless, even if $E(B-V) = 0.83$ 
(Davidge 1998), then $(J-K)_0^{MSTO} = 0.34$, which is still 
significantly larger than the other clusters, although $(J-H)_{0}^{MSTO} = 
0.19$ in this case.

	The MSTO colors of NGC6287 ostensibly suggest that either (1) the 
near-infrared reddening law towards this cluster differs 
significantly from that given by Rieke \& Lebofsky (1985), or 
(2) the spectral-energy distribution (SED) of stars in this cluster 
is somehow peculiar. The $(J-H, H-K)$ two-color diagram (TCD) provides 
additional insight into these possibilities, and the TCD for each cluster is 
shown in Figure 11. In an effort to reduce observational scatter, 
normal points were created by computing the mode 
of the color distribution in $\pm 0.25$ mag (RGB) or $\pm 0.1$ mag (SGB, MSTO) 
bins along the $K$ axis of the CMDs in Figures 8 and 9, and it is the resulting 
$J-H$ and $H-K$ pairs for each $K$ interval which are plotted in Figure 11. 
Also shown in this figure are the loci of metal-poor giant branch and main 
sequence stars listed in Tables 4 and 5 of Davidge \& Harris (1995a). 
Only the AOB data were used to compute normal points, so the most 
evolved giant branch stages are abscent in Figure 11.

	The NGC5272, NGC6205, and NGC6341 measurements fall close to the 
fiducial sequences in Figure 11; however, the NGC6287 points, while 
having $(J-H)_0$ colors similar to those of the other clusters, have relatively 
red $(H-K)_0$ colors. Figure 7 of Davidge (1998) shows a similar tendency for 
NGC6287 stars, although by an amount that is smaller than what is seen 
here. Figure 11 suggests either that the $2\mu$m SEDs of stars 
in NGC6287 may differ from those in other clusters, or that the reddening law 
towards NGC6287 is peculiar. Near-infrared spectra would provide a direct means 
of investigating the SEDs of stars in NGC6287 and provide further insight 
into the placement of NGC6287 data points on the TCD.

\section{ANALYSIS}

	The HB forms an almost vertical sequence in near-infrared CMDs, and so 
the infrared LF of HB stars provides a means of comparing HB content. 
The $K$ HB LFs, normalised according to the total number 
of detected HB stars per cluster, are compared in Figure 12, where $\Delta K$ 
refers to the difference in brightness between each point on the HB and 
the RGB-tip. It is evident that these clusters define a range of HB contents: 
the NGC6205 and NGC6341 HBs are dominated by hot stars 
that are relatively faint in the infrared, while the HB 
stars in NGC5272 and NGC6287 are distributed over a broader range of 
near-infrared brightnesses, and hence temperatures. A Kolmogoroff-Smirnoff 
test indicates that the HB LFs of NGC5272 and NGC6205 differ at the 
99\% confidence level, while the HB LFs of NGC6287 and NGC6341 differ at 
the 98\% confidence level. It is evident that significant information 
concerning HB content can be obtained from near-infrared observations, which 
has important implications for studies of heavily obscured metal-poor clusters. 

	If age is the dominant parameter defining HB content in these 
clusters then NGC6205 should be older than NGC5272, 
while NGC6341 should be older than NGC6287. Is 
there evidence to support this picture? VandenBerg \& Durrell (1990) used the 
brightness difference between the MSTO and RGB-tip to investigate the relative 
ages of clusters in a manner that does not require knowledge of cluster 
distance or reddening. A similar statistic, which we refer to 
as $\Delta^{RGBT}_{SGB}$, that relies on the brightness of the SGB rather than 
the MSTO, can be constructed from the measurements in Table 3, and the results 
are listed in Table 4. The estimated uncertainties in $\Delta^{RGBT}_{SGB}$ are 
$\pm 0.14$ mag, and the entries for NGC5272, NGC6287 and NGC6341 agree 
within these limits. While the $\Delta^{RGBT}_{SGB}$ values for NGC6205 are 
systematically larger than those of the other clusters, the significance of 
this result is not high; indeed, the $\Delta^{RGBT}_{SGB}$ values for NGC5272 
and NGC6205 in $J$ and $H$ differ at only slightly more than the $2\sigma$ 
level.

	The $\pm 0.14$ mag uncertainties in $\Delta^{RGBT}_{SGB}$ correspond to 
a $\pm 4$ Gyr age sensitivity. Hence, we will not consider these results 
further when assessing the relative ages of the clusters, and note that 
extremely high-quality data will be required to obtain interesting age 
constraints from $\Delta^{RGBT}_{SGB}$ measurements. Nevertheless, 
the general agreement between the $\Delta^{RGBT}_{SGB}$ measurements for the 
various clusters lends confidence to the RGB-tip and SGB brightnesses. 

	Another way to estimate relative cluster ages independent oF reddening 
and distance is to register the CMDs using reference markers near the main 
sequence and then compare giant branch locations (e.g. 
VandenBerg, Bolte, \& Stetson 1990, Sarajedini \& Demarque 1990). For the 
current study the $(H, J-H)$ and $(K, J-K)$ CMDs for each cluster 
pair, defined using normal points computed by taking the mode of the color 
distribution within $\pm 0.25$ mag (RGB) and $\pm 0.1$ mag (SGB, MSTO) 
intervals along the brightness axes of Figures 7 and 9, were aligned along the 
horizontal and vertical axes using the MSTO colors and SGB brightnesses listed 
in Table 3, and the results are shown in Figures 13 and 14. 

	The lower giant branches of NGC5272 and NGC6205 overlap when the $(H, 
J-H)$ and $(K, J-K)$ CMDs are registered at the MSTO and SGB. While there are 
differences below the MSTO, these are likely not 
significant as they occur near the faint limit of the data. 
The upper giant branches of NGC5272 and NGC6205, defined using the Cohen {\it 
et al.} (1978) data, are of greater interest, as they diverge on both CMDs. 
There is no hint of this behaviour when only the AOB data are compared, and we 
suspect that the disagreement is spurious, arising from minor 
differences in the photometric calibration of the two NGC5272 datasets. 
Indeed, if the Cohen {\it et al.} (1978) NGC5272 measurements in Figure 14 are 
shifted by $\sim 0.04$ mag towards smaller $\Delta(J-K)$ values then they 
give a better match to the AOB data for this cluster as well as 
to the NGC6205 upper giant branch. 

	The lower giant branches of NGC6287 and NGC6341 also overlap when the 
$(H, J-H)$ CMDs of these clusters are registered at the MSTO and SGB. However, 
the cluster sequences show significant differences at the bright end, in that 
NGC6287 has a steeper upper giant branch than NGC6341; 
Davidge (1998) found that the upper giant branch of NGC6287 is 
steeper than expected from theoretical models. Further 
differences between the CMDs of NGC6287 and NGC6341 are apparent in the right 
hand panel of Figure 14, where the giant branch of NGC6287 
falls consistently to the left of the NGC6341 sequence.

	NGC6287 is not the only inner spheroid cluster to show peculiar 
giant branch properties. Figure 4 of Minniti, Olszewski, \& Rieke (1995) 
compares the $(K, J-K)$ giant branch loci of a number of inner spheroid 
clusters, and there is little or no evidence for correlations between mean 
metallicity, giant branch color, and/or slope. Moreover, the $(K, J-K)$ CMDs of 
the inner spheroid clusters M22 (Davidge \& Harris 1995b) and M28 (Davidge {\it 
et al.} 1996) show peculiar giant branch morphologies when compared with 
outer halo clusters.

	The error bars in the lower right hand corner of each panel in Figures 
13 and 14 show the range in relative RGB positions expected for 
$\pm 2$ Gyr age differences, as predicted from the [Fe/H] = --2 Bergbusch 
\& VandenBerg (1992) isochrones, which we transformed onto the near-infrared 
observational plane using the procedure described by Davidge \& Harris (1995a). 
The lower giant branches of NGC5272 and NGC6205 
in Figures 13 suggest that the ages of these clusters 
differ by no more than $\pm 1$ Gyr. However, while the 
lower giant branches of NGC6287 and NGC6341 in Figure 13 are 
suggestive of a difference in age not exceeding $\pm 2$ Gyr, the 
comparison in the right hand panel of Figure 14 is suggestive of an age 
difference approaching 4 Gyr, in the sense that NGC6287 is older. 

	The LFs of globular clusters are sensitive to age variations at the 
brightness of the SGB (e.g. Bergbusch \& VandenBerg 1992; Jimenez \& Padoan 
1996), and hence provide an independent means of checking relative ages 
deduced from CMDs. While dynamical effects will alter the mass function, and 
thereby complicate cluster-to-cluster comparisons, the LF 
only becomes sensitive to these effects below the main sequence. 

	The $J$, $H$, and $K$ LFs of each cluster pair, computed with 0.2 mag 
bins and excluding HB and AGB stars identified from the CMDs, are 
compared in Figures 15, 16, and 17. The LFs have been normalized along the 
vertical axis over a 3 mag interval on the upper giant branch, 
while the horizontal axis shows brightness measured with respect to the 
RGB-tip. The $J$ LFs are not suitable for age comparisons, as incompleteness 
becomes signicant on the SGB. However, the $H$ and $K$ LFs extend 0.5 -- 1.0 
mag fainter than the MSTO, and it is evident that the LFs of each cluster pair 
are not significantly different on the SGB. The effect of a $\pm 2$ Gyr age 
variation near the faint end of the SGB, derived from the Bergbusch \& 
VandenBerg (1992) models, is shown in each Figure, and the expected 
deviation is comparable to the uncertainties in each bin. Hence, 
the $H$ and $K$ LFs indicate that NGC5272 and 
NGC6205 are coeval, as are NGC6287 and NGC6341, to within $\pm 2$ Gyr.

\section{SUMMARY \& DISCUSSION}

	$JHK$ images with angular resolutions approaching the diffraction 
limit of the 3.6 metre CFHT have been obtained of the central regions of four 
metal-poor globular clusters: NGC5272, NGC6205, NGC6287, and NGC6341. The 
clusters were paired according to metallicity and distance to facilitate 
differential comparisons of cluster properties, and photometric measurements 
were made using the PSF-fitting routine in ALLSTAR. The crowded nature of the 
cluster fields necessitated the use of a single PSF 
for each cluster$+$filter combination, and it is expected that PSF 
variations will introduce systematic errors of roughly 0.01 mag in the 
brightnesses of stars more than 8.5 arcsec from the reference star, which 
account for roughly 80\% of stars in the images. These 
errors may climb to a few hundredths of a mag for stars within 8.5 arcsec of 
the reference star. Comparisons with published observations indicate that the 
photometric calibration is valid to within a few hundredths of a magnitude.

	The HB contents of each cluster pair were investigated using the $K$ LF 
of HB stars. The `classical' second-parameter pair NGC5272 and NGC6205 provides 
an important test case to determine if differences between HB contents can be 
detected in the infrared, and the $K$ HB LFs of these clusters
differ in a way that is consistent with what is seen at optical 
wavelengths. The temperature distribution of HB stars in NGC6287 and NGC6341 
are also found to be significantly different.

	Relative cluster ages have been determined using two independent 
methods: (1) the color difference between the MSTO and the giant 
branch, measured from near-infrared CMDs, and (2) 
the near-infrared LFs of SGB stars. A third possible age indicator, the 
brightness difference between the RGB-tip and a point on the SGB, was 
too insensitive to age to be of interest with the current data. 

	The near-infrared CMDs of NGC5272 and NGC6205, when registered near 
the MSTO and SGB, indicate that the ages of these clusters 
agree to within $\pm 1$ Gyr, while the SGB LFs 
are suggestive of ages that agree to within $\pm 2$ Gyr. These data 
thus do not support the notion that age is the second parameter (e.g. Salaris 
\& Weiss 1997, Stetson, VandenBerg, \& Bolte 1996, but see also Chaboyer {\it 
et al.} 1996 and Lee, Demarque, \& Zinn 1994). Grundahl (1998) has obtained 
high-precision ground-based observations of NGC5272 and NGC6205 in the 
Stromgren filter system, and his data are also consistent with little or no 
difference in age.

	The problems arising from field star contamination and differential 
reddening that frustrate efforts to study the outer regions of NGC6287 (e.g. 
Davidge 1998, Stetson \& West 1994) are largely avoided when the high-density 
central area of the cluster is studied, and the current data provide 
insight into the relative ages of NGC6287 and NGC6341. The 
$(H, J-H)$ CMDs and near-infrared LFs suggest that the ages of these clusters 
agree to within $\pm 2$ Gyr, and thus do not support the 
hypothesis that the inner regions of the Galaxy formed significantly before 
the outer halo, contrary to what might be infered from the properties of 
RR Lyraes in Baade's Window (Lee 1992a). On the other hand, the $(K, J-K)$ CMDs 
of NGC6287 and NGC6341 ostensibly suggest that the former is 4 Gyr older than 
the latter. A similar difference in age was found when the $(K, J-K)$ CMD of 
the [Fe/H] $= -1.4$ globular cluster M28 (R$_{GC} 
\sim 2.5$ kpc) was compared with CMDs of clusters having 
R$_{GC} \geq 3$ kpc (Davidge, C\^{o}t\'{e}, \& Harris 1996; Davidge \& 
Harris 1997). NGC6287 is thus the second inner spheroid cluster to have 
an extremely old age infered from the $(K, J-K)$ CMD.
However, the significance of this last result is not clear, as 
the location of NGC6287 stars on the near-infrared TCD 
suggest that the $2\mu$m photometric properties of metal-poor stars in this 
cluster differ from those of outer halo objects. 
This brings into question the validity of 
$(K, J-K)$ CMD comparisons between inner and outer halo clusters. 

	The relative ages of NGC6287 and NGC6341 notwithstanding, NGC6287 
appears to be a unique object. With a plunging orbit (van den Bergh 1993), and 
a broadly populated HB (\S 4), NGC6287 has the characteristics of van den 
Bergh's (1993) $\alpha$ population of globular clusters, the other members of 
which have R$_{GC} \geq 8$ kpc. The discovery of HB stars spanning a range of 
temperatures in NGC6287 also challenges Lee's (1992b) finding that the HB 
content of inner spheroid globular clusters can be characterised by a single 
parameter -- metallicity; Rich et al. (1997) reach a similar conclusion for 
metal-rich globular clusters. Deep photometric surveys of other metal-poor 
clusters in the inner Galaxy will reveal if the HB properties of NGC6287 are 
truely unique, or representative of a still largely unstudied population of 
inner spheroid clusters.

	If NGC6287 is a bona fide member of the 
inner spheroid then it is a remnant of what was once a much more populous 
cluster system, the members of which, especially those that had low masses, low 
concentrations, and circular orbits, were disrupted by dynamical interactions 
(e.g. Murali \& Weinberg 1997, Vesperini 1997 and references therein). 
If this is the case then the initial properties of NGC6287 were 
undoubtedly very different from what are observed now. For example, selective 
pruning of low mass stars has likely occured, and the current mass may be 
more than $30\%$ lower than the initial value (Murali \& Weinberg 1997). 
Studies of the main sequence mass function, especially near the center of 
NGC6287, where the signatures of mass segregation are conspicuous 
(e.g. Weinberg 1994, Pryor, Smith, \& McClure 1986), 
will provide clues into the dynamical history of this cluster.

	The discovery of the Sgr dwarf spheroidal 
(Ibata, Gilmore, \& Irwin 1994), which has an entourage of four globular 
clusters (Ibata {\it et al.} 1994; Da Costa \& Armandroff 1995), suggests that 
accretion events may play a significant role in the evolution of galaxies, 
and that the present-day sample of globular clusters 
in the inner Galaxy may contain interlopers that formed in 
external systems. In fact, the radius and metallicity of NGC6287 are not 
inconsistent with a dwarf spheroidal origin. Indeed, van den 
Bergh (1994) finds that the globular clusters in the Fornax dwarf spheroidal 
have radii comparable to those of globular clusters in the inner halo, so the 
compact size of NGC6287 is not an unambiguous indicator of an 
inner spheroid origin. Moreover, the metallicities of globular clusters in 
the dwarf satellite companions of the Galaxy (Da Costa \& Armandroff 1995) and 
M31 (Da Costa \& Mould 1988) are typically (but, as demonstrated by Terzan 7, 
not exclusively) very metal-poor, like NGC6287. Therefore, while the 
observational characteristics of NGC6287 are ostensibly consistent 
with membership in the inner spheroid (van den Bergh 1993), the 
evidence is not ironclad. 

\vspace{1.0cm}
	Sincere thanks are extended to Sidney van den Bergh and Frank 
Grundahl for providing comments on an earlier draft of this paper. An anonymous 
referee also made comments that greatly improved the paper.


\clearpage

\begin{table*}
\begin{center}
\begin{tabular}{lcccrcc}
\tableline\tableline
NGC & [Fe/H] & $E(B-V)$ & V$^{HB}_0$ & R$_{GC}$ & $\mu^{V}_0$ & M$_V$ \\
 & & & & (kpc) & mag/arcsec$^2$ & \\
\tableline
5272 & --1.57 & 0.01 & 15.62 & 11.9 & 16.31 & --8.85 \\
6205 & --1.54 & 0.02 & 14.84 & 8.3 & 16.74 & --8.50 \\
6287 & --2.05 & 0.59 & 15.17 & 1.6 & 16.50 & --7.11 \\
6341 & --2.29 & 0.02 & 15.04 & 9.5 & 15.52 & --8.15 \\
\tableline
\end{tabular}
\end{center}
\caption{Cluster Properties}
\end{table*}


\clearpage

\begin{table*}
\begin{center}
\begin{tabular}{lccccc}
\tableline\tableline
NGC & $\alpha$ & $\delta$ & Filter & Exposure Time & FWHM \\
 & E2000 & E2000 & & (sec) & (arcsec) \\
\tableline
5272 & 13:42:11.2 & $+$28:32:22 & $J$ & $8 \times 30$ & 0.31 \\
 & & & $H$ & $8 \times 30$ & 0.14 \\
 & & & $K$ & $8 \times 30$ & 0.14 \\
 & & & & & \\
6205 & 16:41:39.9 & $+$36:27:36 & $J$ & $4 \times 90$ & 0.29 \\
 & & & $H$ & $4 \times 90$ & 0.26 \\
 & & & $K$ & $4 \times 90$ & 0.22 \\
 & & & & & \\
6287 & 17:05:09.3 & $-$22:42:27 & $J$ & $4 \times 60$ & 0.24 \\
 & & & $H$ & $4 \times 60$ & 0.15 \\
 & & & $K$ & $4 \times 60$ & 0.17 \\
 & & & & & \\
6341 & 17:15:34.4 & $+$43:11:01 & $J$ & $4 \times 30$ & 0.22 \\
 & & & $H$ & $4 \times 20$ & 0.14 \\
 & & & $K$ & $4 \times 15$ & 0.15 \\
\tableline
\end{tabular}
\end{center}
\caption{Summary of Observations}
\end{table*}


\clearpage

\begin{table*}
\begin{center}
\begin{tabular}{lcccccccc}
\tableline\tableline
NGC & $(J-H)_{MSTO}$ & $(J-K)_{MSTO}$ & $J_{SGB}$ & $H_{SGB}$ & $K_{SGB}$ & $J_{RGBT}$ & $H_{RGBT}$ & $K_{RGBT}$ \\
\tableline
5272 & 0.239 & 0.234 & 16.56 & 16.22 & 16.63 & 10.08 & 9.29 & 9.16 \\
6205 & 0.225 & 0.290 & 16.28 & 16.06 & 16.17 & 9.32 & 8.60 & 8.49 \\
6287 & 0.461 & 0.774 & 16.62 & 16.33 & 15.98 & 10.18 & 9.25 & 9.00 \\
6341 & 0.210 & 0.230 & 16.10 & 15.86 & 16.09 & 9.69 & 9.05 & 8.93 \\
\tableline
\end{tabular}
\end{center}
\tablecomments{The errors in the MSTO colors are $\pm 0.03$ mag. 
The estimated uncertainty in the SGB and RGB-tip brightnesses is $\pm 
0.1$ mag.}
\caption{The colors and brightnesses of key features on the CMDs}
\end{table*}


\clearpage

\begin{table*}
\begin{center}
\begin{tabular}{lccccc}
\tableline\tableline
NGC & $(J-H)_{0}^{MSTO}$ & $(J-K)_{0}^{MSTO}$ & $\Delta J_{SGB}^{RGBT}$ & $\Delta H_{SGB}^{RGBT}$ & $\Delta K_{SGB}^{RGBT}$ \\
\tableline

5272 & 0.24 & 0.23 & --6.48 & --6.93 & --7.47 \\
6205 & 0.22 & 0.28 & --6.96 & --7.38 & --7.68 \\
6287 & 0.27 & 0.47 & --6.43 & --7.08 & --6.98 \\
6341 & 0.20 & 0.22 & --6.41 & --6.81 & --7.16 \\
\tableline
\end{tabular}
\end{center}
\tablecomments{The errors in the de-reddened MSTO colors are $\pm 0.03$ mag, 
with the exception of NGC6287, for which the reddening is highly uncertain. The 
estimated uncertainty in individual $\Delta^{RGBT}_{SGB}$ measurements is $\pm 
0.14$ mag.}
\caption{MSTO colors and brightness differences between the RGB-tip and SGB.}
\end{table*}


\clearpage

\begin{center}
FIGURE CAPTIONS
\end{center}

\figcaption
[davidge.fig1.gif]
{The final $K$ image of the NGC5272 field. The area covered is 
roughly $34 \times 34$ arcsec. North is at the top, while East is to the left.}

\figcaption
[davidge.fig2.gif]
{Same as Figure 1, but for NGC6205.}

\figcaption
[davidge.fig3.gif]
{Same as Figure 1, but for NGC6287.}

\figcaption
[davidge.fig4.gif]
{Same as Figure 1, but for NGC6341.}

\figcaption
[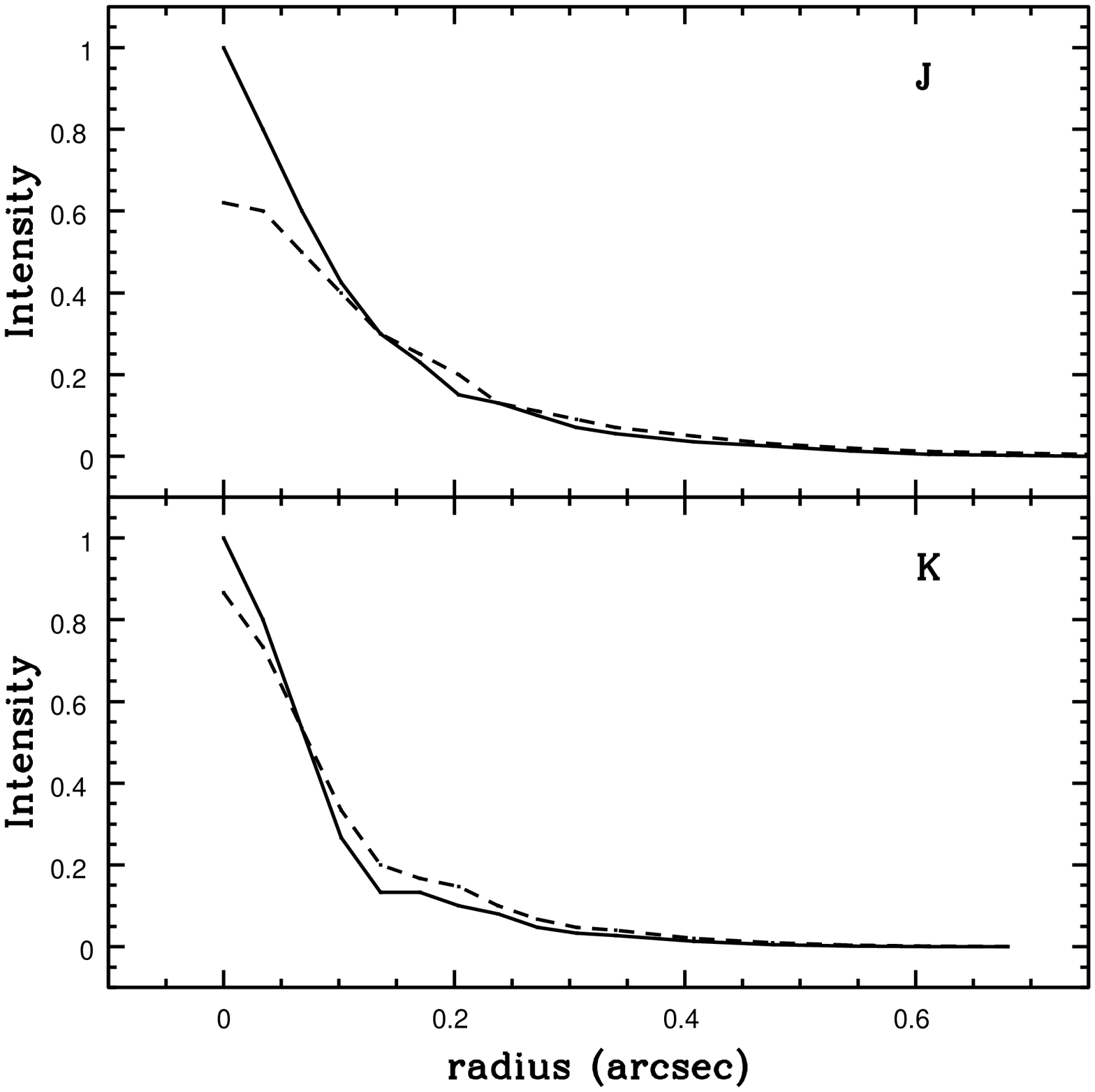]
{$J$ and $K$ PSFs in the NGC6287 field. The solid lines show the 
PSF within 8.5 arcsec of the AO reference star, while the dashed lines show the 
PSF between 8.5 and 17 arcsec from this star. The effects of 
anisoplanaticism are greatest in $J$}

\figcaption
[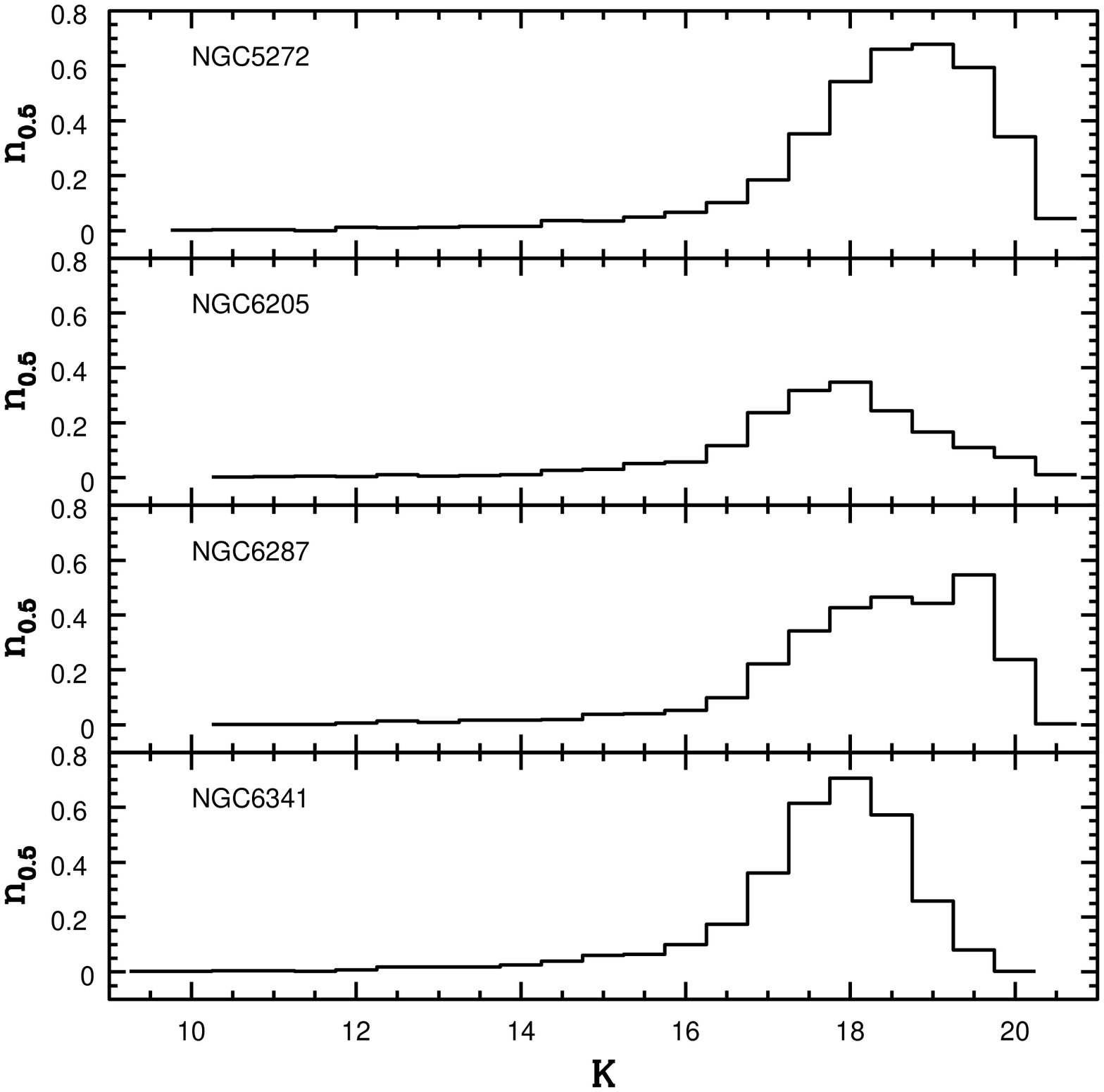]
{The $K$ LFs of the four clusters. n$_{0.5}$ is the number of 
stars per 0.5 mag interval per square arcsec.}

\figcaption
[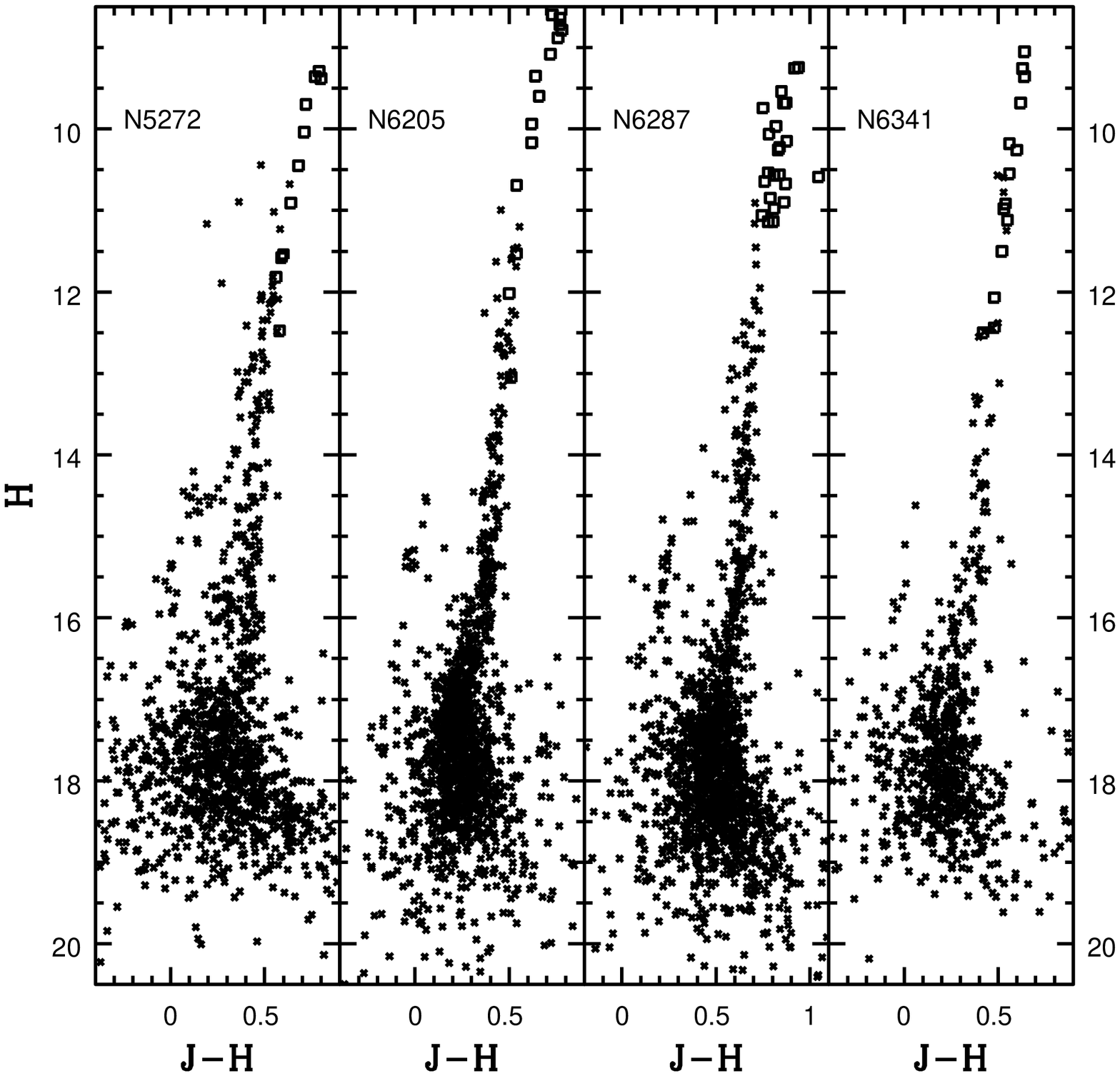]
{The $(H, J-H)$ CMDs of NGC5272, NGC6205, NGC6287, and NGC6341. The open 
squares show the photometric measurements of bright giants published by Cohen 
{\it et al.} (1978) for NGC5272, NGC6205, and NGC6341, and Davidge (1998) for 
NGC6287. Note the good agreement between the published measurements and the 
AO-corrected data.}

\figcaption
[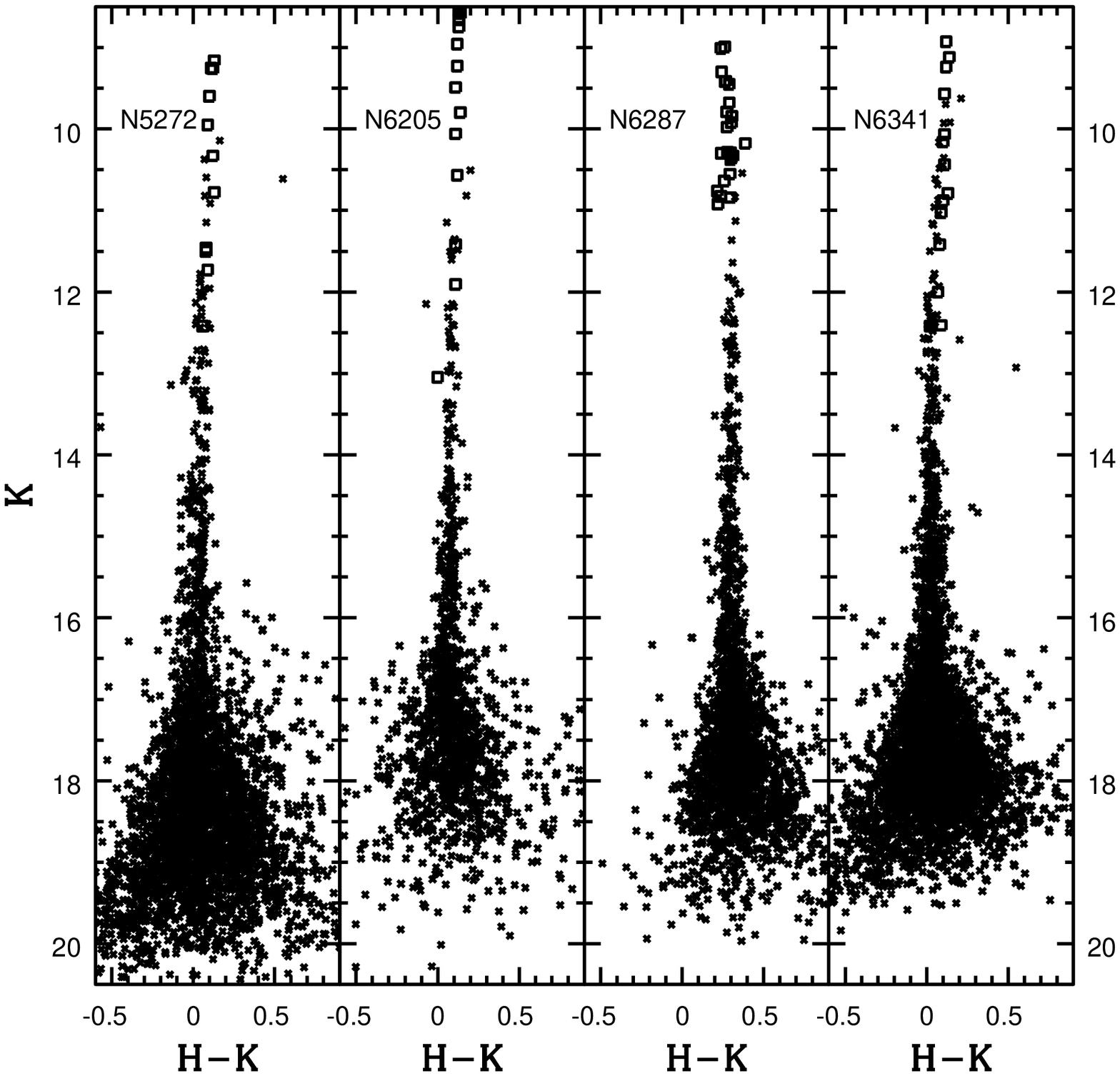]
{The $(K, H-K)$ CMDs of NGC5272, NGC6205, NGC6287, and NGC6341. The open 
squares show the photometric measurements of bright giants published by Cohen 
{\it et al.} (1978) for NGC5272, NGC6205, and NGC6341, and Davidge (1998) for 
NGC6287. Note the good agreement between the published measurements and the 
AO-corrected data.}

\figcaption
[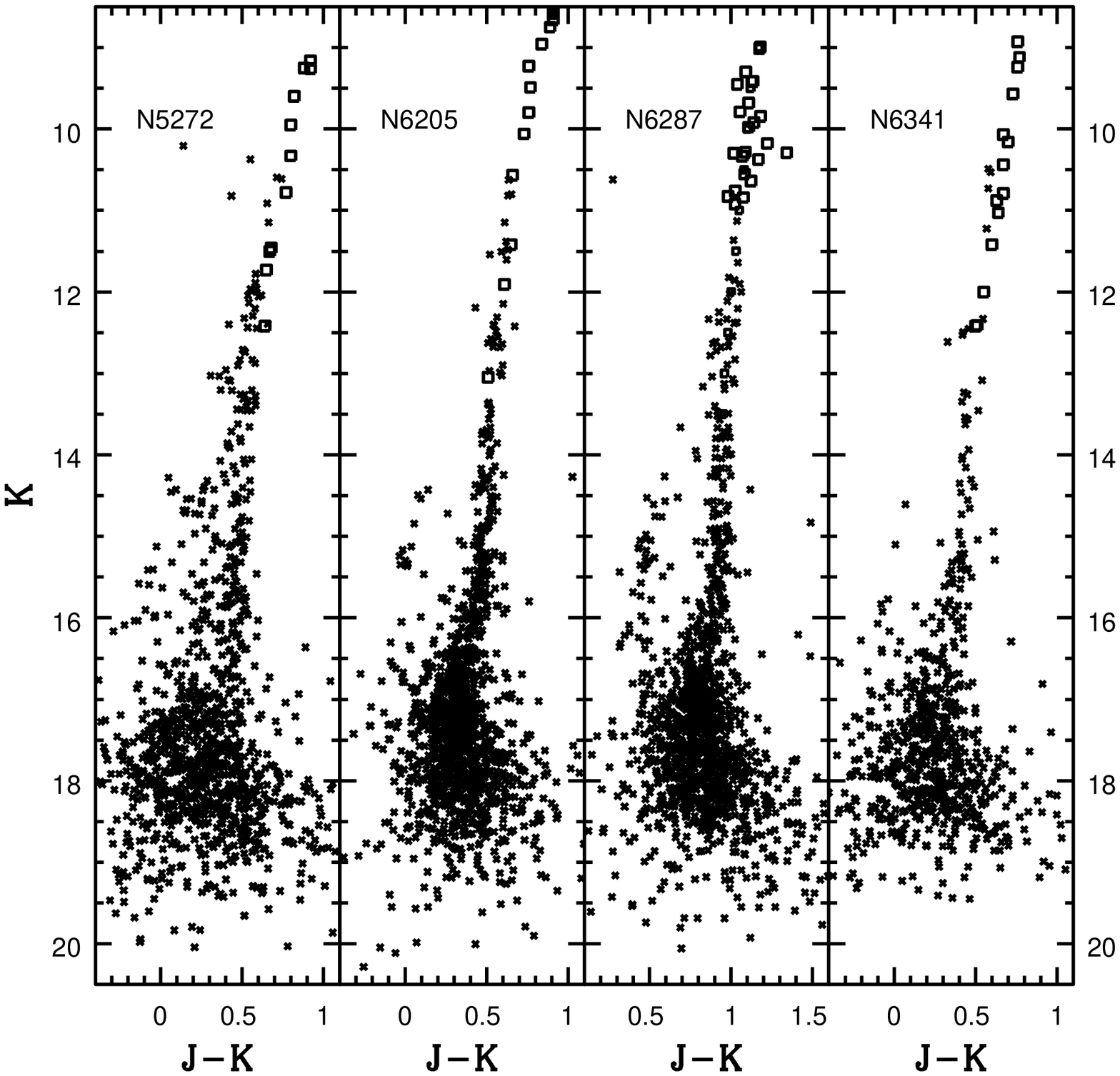]
{The $(K, J-K)$ CMDs of NGC5272, NGC6205, NGC6287, and NGC6341. The open 
squares show the photometric measurements of bright giants published by Cohen 
{\it et al.} (1978) for NGC5272, NGC6205, and NGC6341, and Davidge (1998) for 
NGC6287. Note the good agreement between the published measurements and the 
AO-corrected data.}

\figcaption
[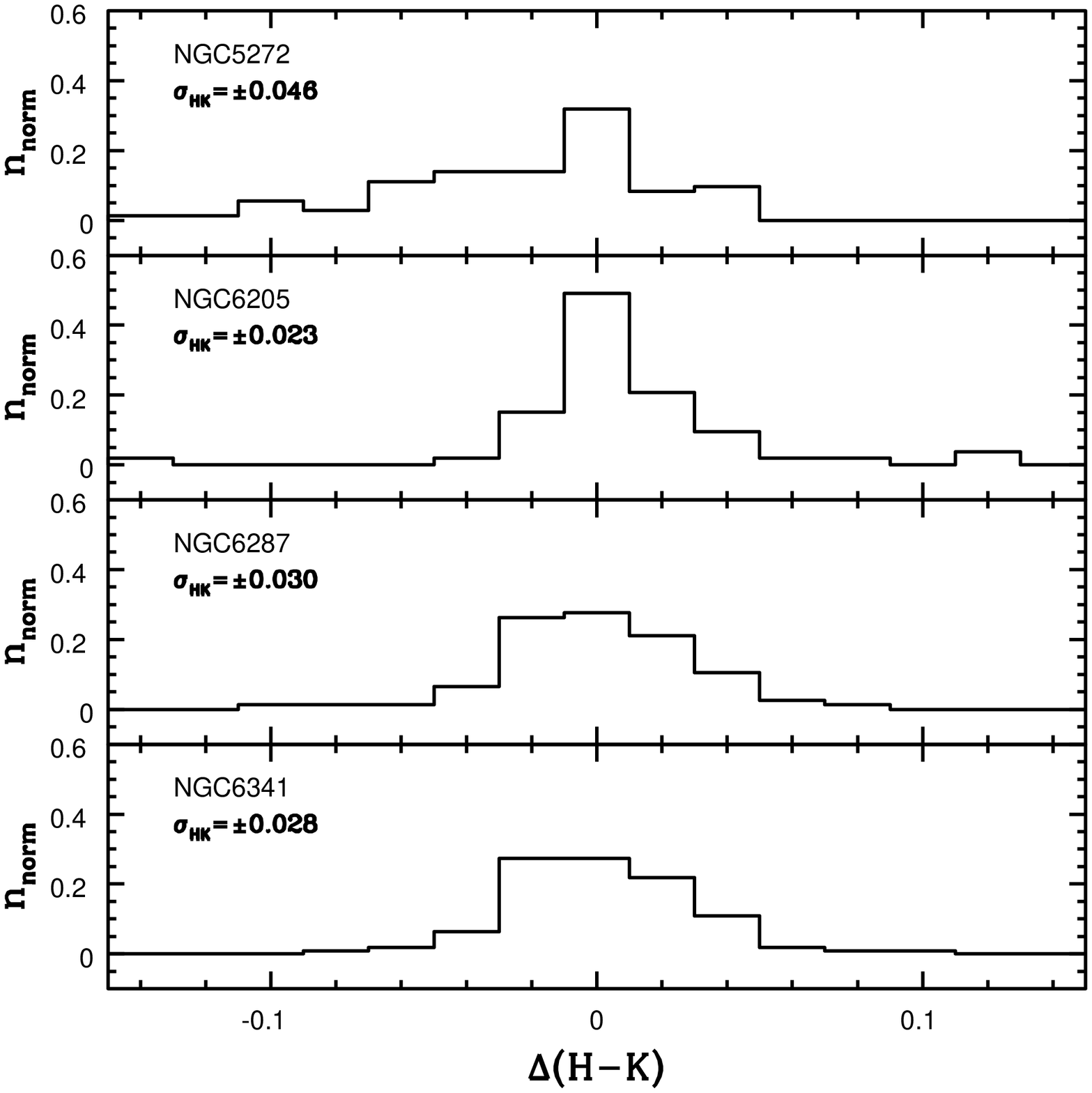]
{The $H-K$ color distributions for stars with $K$ between 12 and 14.5. 
$\Delta$(H--K) is the difference between the observed H--K and the peak value 
in each distribution. n$_{norm}$ is the number of stars per 0.02 mag 
$\Delta$(H--K) interval divided by the total number of stars in the 
distribution. The standard deviation of each distribution, $\sigma _{HK}$, is 
also shown.}

\figcaption
[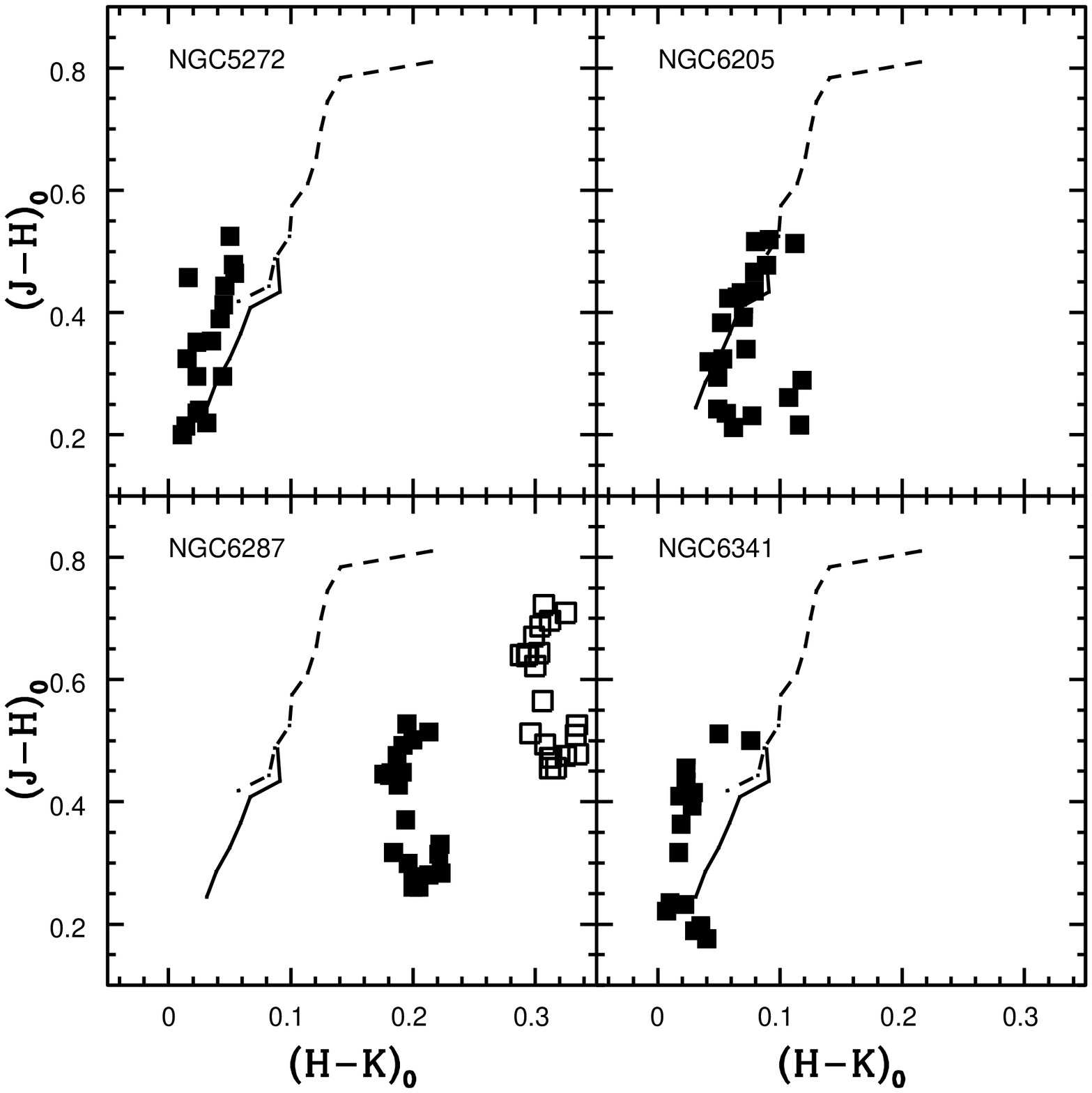]
{The $(J-H, H-K)$ two-color diagrams for NGC5272, NGC6205, 
NGC6287, and NGC6341. The points plotted in this figure are 
the modes of the color distribution in $\pm 0.25$ mag (RGB) or 
$\pm 0.1$ mag (SGB, MSTO) wide bins along the $K$ axis of the CMDs in Figures 
8 and 9; each point in this figure corresponds to the $J-H$ and 
$H-K$ normal points for a given $K$ interval. These normal points have 
been de-reddened using the E(B--V) values in Table 1 according to the 
Rieke \& Lebofsky (1985) reddening law. Also shown are the loci defined 
by metal-poor dwarfs (solid line) and giants (dashed line), as listed in Tables 
4 and 5 of Davidge \& Harris (1995a). The open squares show the unreddened 
NGC6287 normal points.}

\figcaption
[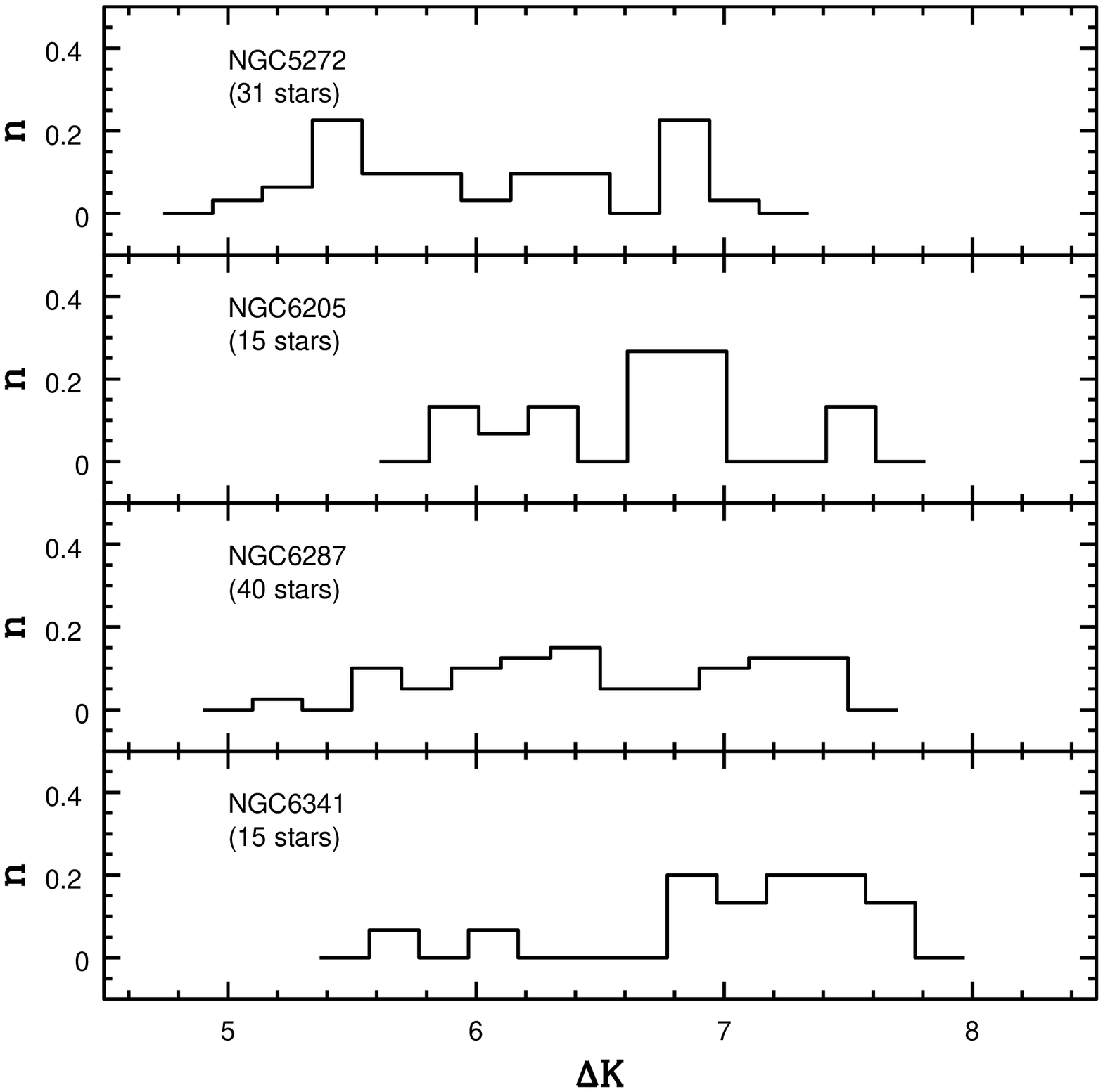]
{The $K$ HB LFs, based on stars identified from 
the $(K, J-K)$ CMDs. $n$ is the number of stars per 0.2 mag interval normalised 
according to the total number of HB stars in each field, which is the 
number shown in brackets under each cluster name. $\Delta$K is the mag 
difference between each point on the HB and the RGB-tip.}

\figcaption
[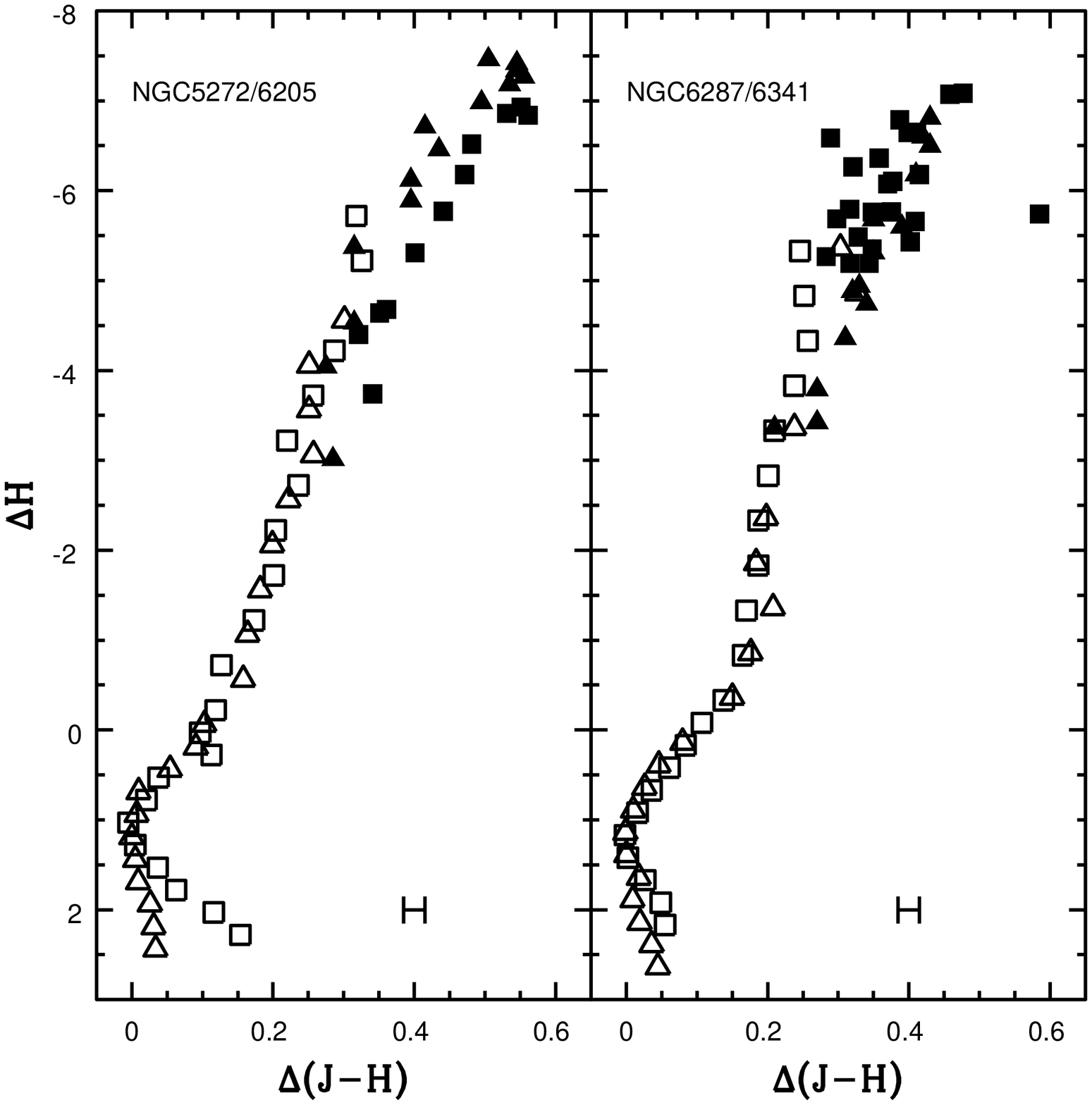]
{The $(H, J-H)$ CMDs for each cluster pair, registered along the 
horizontal and vertical axes based on the color of the MSTO and the brightness 
of a point on the SGB 0.1 mag redward of the MSTO. $\Delta$(J--H) is the 
observed color minus the color of the MSTO, while $\Delta$H is the observed 
brightness minus the brightness of the SGB point. The open squares show 
the NGC5272 (left hand panel) and NGC6287 (right hand panel) normal points. 
The open triangles are normal points for NGC6205 (left hand panel) and 
NGC6341 (right hand panel). The filled symbols are observations of individual 
bright stars in these clusters from Davidge (1998) and Cohen {\it et al.} 
(1978). The range in RGB positions expected for a $\pm 2$ Gyr age difference 
is shown in the lower right hand corner of each panel.}

\figcaption
[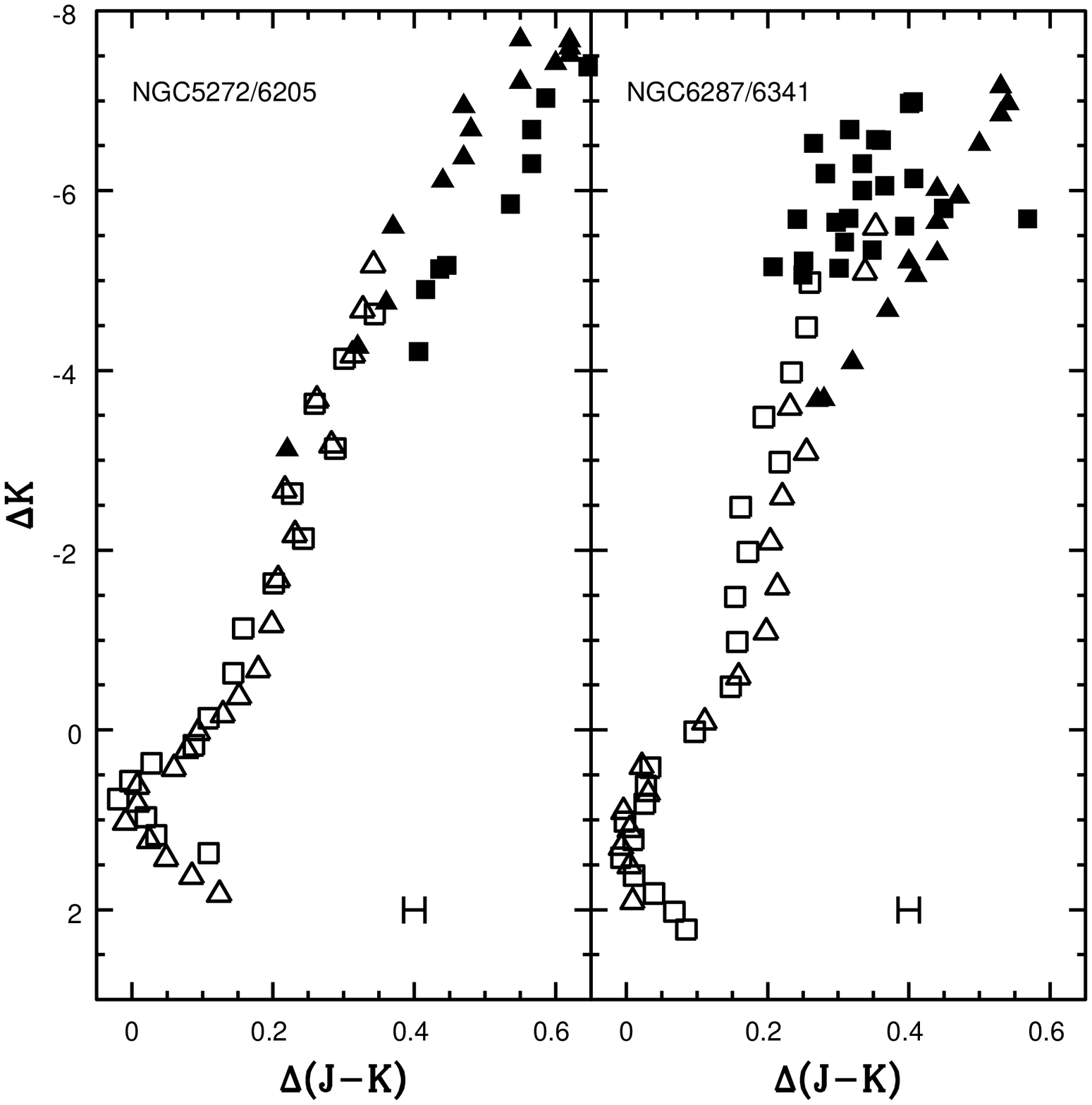]
{Same as Figure 12, but for the $(K, J-K)$ CMDs.}

\figcaption
[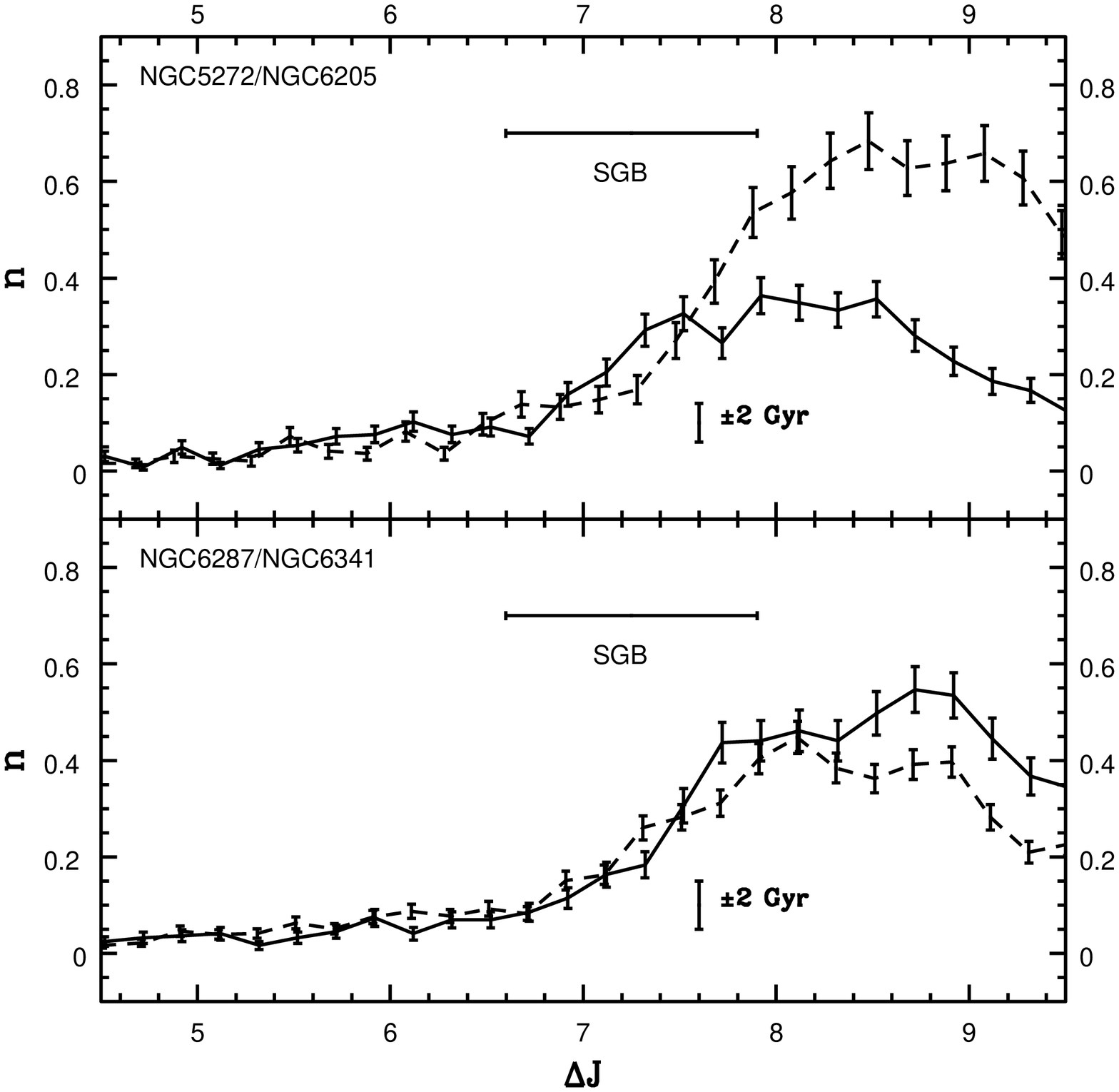]
{The $J$ LFs of the cluster pairs, excluding HB and AGB stars which 
were identified from the CMDs. The LFs for NGC5272 (solid line) 
and NGC6205 (dashed line) are shown in the top panel, while the LFs for 
NGC6287 (solid line) and NGC6341 (dashed line) are shown in the lower panel. 
$\Delta$J is the brightness measured with respect to the RGB-tip, and the 
LFs have been normalised based on the number of stars detected between 
$\Delta$J = 3 and 7. The error bars show the uncertainty in each 0.2 mag bin 
predicted from counting statistics. The brightness interval corresponding to 
the SGB is indicated in each panel. The variation in LF position 
near the faint limit of the SGB predicted for a $\pm 2$ Gyr age variation 
is also shown.}

\figcaption
[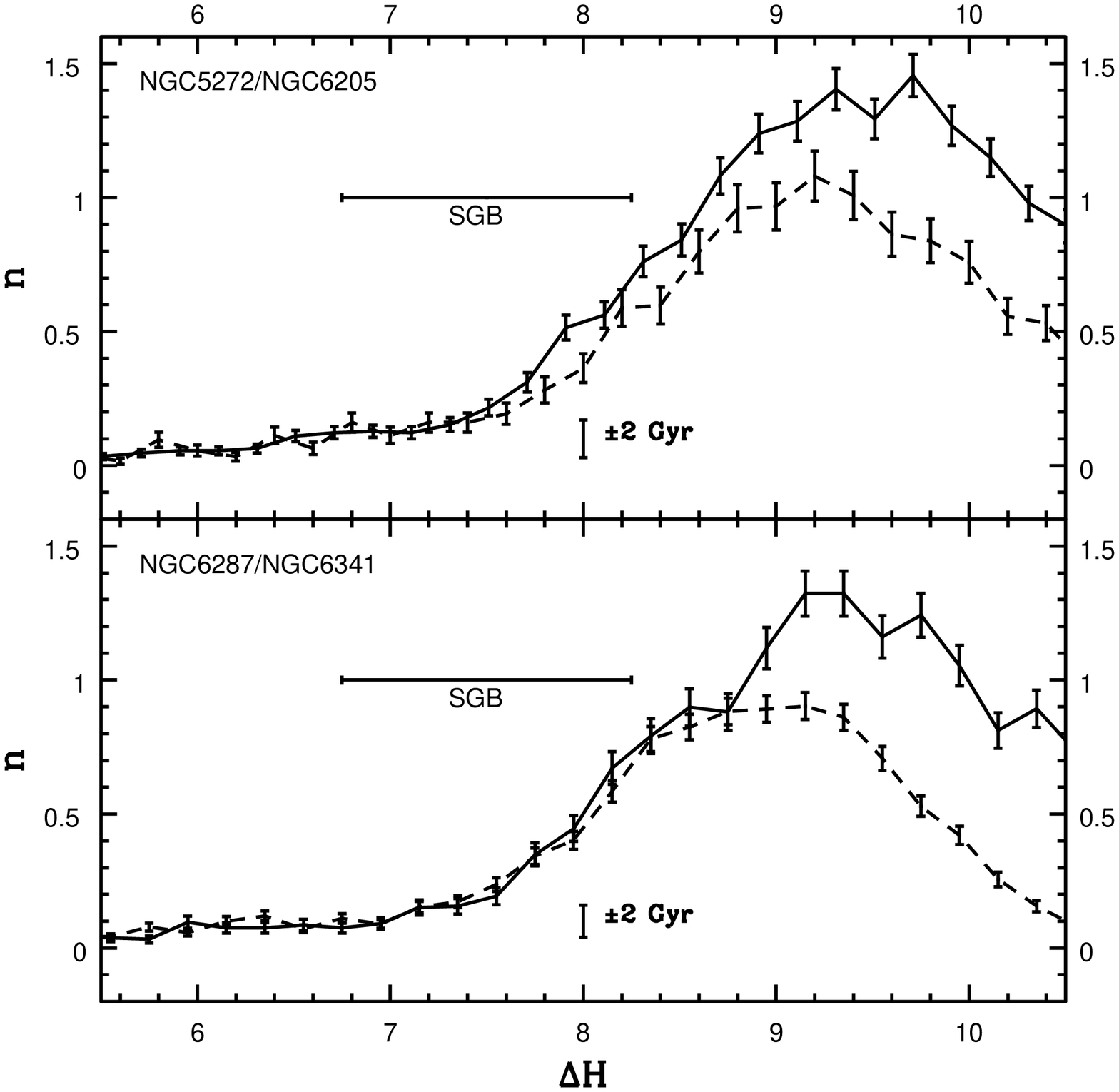]
{Same as Figure 15, but for $H$. The LFs have been normalised 
based on the number of stars detected between $\Delta$H = 3 and 7.}

\figcaption
[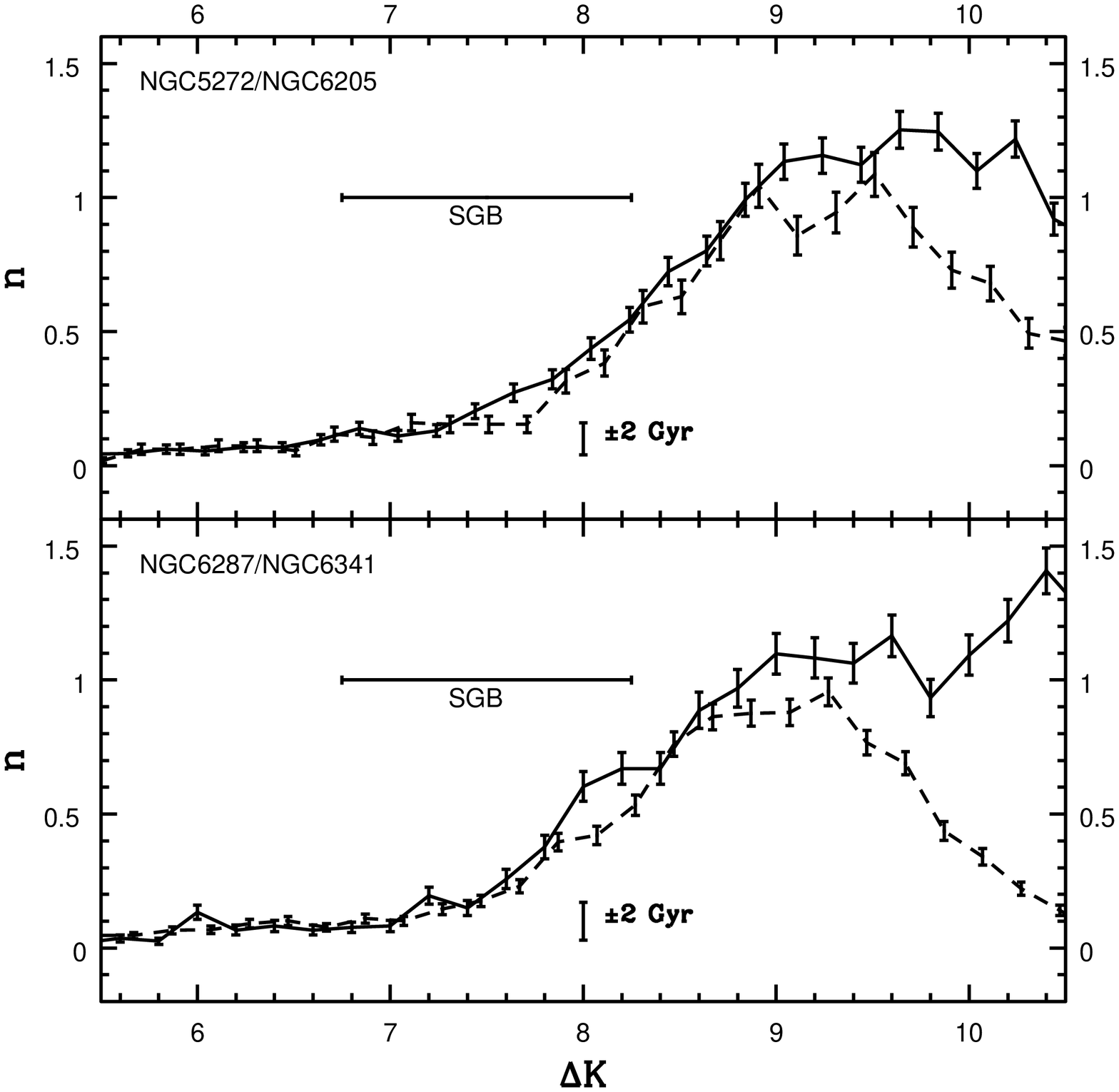]
{Same as Figure 15, but for $K$. The LFs have been normalised 
based on the number of stars detected between $\Delta$K = 3 and 7.}
\end{document}